\documentclass[10pt,conference,letterpaper]{IEEEtran}
\IEEEoverridecommandlockouts
\usepackage{cite}
\usepackage{amsmath,amssymb,amsfonts}
\usepackage{algorithmic}
\usepackage[
  separate-uncertainty = true,
  multi-part-units = repeat
]{siunitx}
\usepackage{graphicx}
\usepackage{textcomp}
\usepackage{xcolor}
\usepackage{multirow}
\def\BibTeX{{\rm B\kern-.05em{\sc i\kern-.025em b}\kern-.08em
    T\kern-.1667em\lower.7ex\hbox{E}\kern-.125emX}}
\usepackage{amssymb}

\begin{document}

\title{Optimised Convolutional Neural Networks for Heart Rate Estimation and Human Activity Recognition in Wrist Worn Sensing Applications\\
\thanks{This work is part-funded by Science Foundation Ireland under grant numbers 17/RC-PhD/3482 SFI/12/RC/2289 and SAP SE}
}

\author{
    \IEEEauthorblockN{Eoin Brophy\IEEEauthorrefmark{1}, Willie Muehlhausen\IEEEauthorrefmark{2}, Alan F. Smeaton\IEEEauthorrefmark{2}, Tom{\'a}s E. Ward\IEEEauthorrefmark{2}}
    \IEEEauthorblockA{\IEEEauthorrefmark{1}Infant Research Centre, University College Cork}
    \IEEEauthorblockA{\IEEEauthorrefmark{2}Insight Centre for Data Analytics, Dublin City University
    \\\{eoin.brophy, willie.muehlhausen, tomas.ward, \\alan.smeaton\}@insight-centre.org}
}

\maketitle
\begin{abstract}
Wrist-worn smart devices are providing increased insights into human health, behaviour and performance through sophisticated analytics. However, battery life, device cost and sensor performance in the face of movement-related artefact present challenges which must be further addressed to see effective applications and wider adoption through commoditisation of the technology. We address these challenges by demonstrating, through using a simple optical measurement, photoplethysmography (PPG) used conventionally for heart rate detection in wrist-worn sensors, that we can provide improved heart rate and human activity recognition (HAR) simultaneously at low sample rates, without an inertial measurement unit. This simplifies hardware design and reduces costs and power budgets. We apply two deep learning pipelines, one for human activity recognition and one for heart rate estimation. HAR is achieved through the application of a visual classification approach, capable of robust performance at low sample rates. Here, transfer learning is leveraged to retrain a convolutional neural network (CNN) to distinguish characteristics of the PPG during different human activities. For heart rate estimation we use a CNN adopted for regression which maps noisy optical signals to heart rate estimates. In both cases, comparisons are made with leading conventional approaches. Our results demonstrate a low sampling frequency can achieve good performance without significant degradation of accuracy. 5 Hz and 10 Hz were shown to have 80.2\% and 83.0\% classification accuracy for HAR respectively.  These same sampling frequencies also yielded a robust heart rate estimation which was comparative with that achieved at the more energy-intensive rate of 256 Hz. 

\end{abstract}


\section{Introduction}
Photoplethysmography (PPG) is an optical technique commonly employed in wearables and other medical devices to measure volume changes of blood in the microvascular tissue during the cardiac cycle. Light becomes reflected and absorbed at different rates during this cycle and the reflected light is read by a photo-sensor to detect these changes. The output from this sensor is then processed so a valid heart rate estimation can be determined.  

Heart rate can be measured at multiple sites on the body using PPG including, but not limited to; ear, forehead, fingertip, ankle and wrist. In the context of personalised health and fitness monitoring using wearables, the wrist is the most frequently used location for photoplethysmographic heart rate monitoring. Accuracies of consumer-grade wearables, for the most part, are acceptable but are prone to errors during daily activities \cite{Nelson2019}. The difficulties associated with correctly estimating heart rate arise mostly in obtaining a strong physiological reading from the sensors. Often the signals read from the PPG modules are heavily corrupted with motion artefacts and the movement of the limbs is a major contributor to this introduced artefact. Retrieval of a clean PPG signal from a heavily corrupted signal can be achieved by applying filtering techniques including adaptive methods based on a measure of the artefact sourced from an accelerometer-based measurement \cite{Allen2007}.

We have shown previously that human activity recognition (HAR) can be performed on optical signals, taking advantage of the artefact present in the signal \cite{Brophy2018}. In this study we take this a step further, exploring to what extent HAR is sufficiently accurate when decreasing the sampling frequency and investigating whether we can obtain a valid heart rate estimation without further on-board filtering of the PPG signal thus reducing computing requirements. 

The battery life of smartwatches and fitness trackers vary greatly depending on the features and functionality available on-board the wearable. The Apple Watch Series 5, which is more of a lifestyle and fitness tracker, can run for a period of up to 18 hours whereas the Fitbit Charge 3 fitness tracker can go for up to 7 days on a single charge. Continuous activity and heart rate monitoring speed up the depletion of the battery of most wearables. Gathering and processing of simultaneous sensor data can further increase the power consumption of the devices. Without explicitly stating the sampling frequency, Apple state that their heart rate monitor Light Emitting Diodes (LEDs) blink “hundreds of times per second” \cite{AppleHT}.

Capitalising on recent advancements in machine learning could pave the way for the simplification of wearables, allowing for a reduction in power requirements and subsequently smaller and lower-cost devices. The work described in this paper is part of a larger-scoped effort to develop easily deployed artificial intelligence which can be used and interpreted by end-users who do not have deep levels of signal processing expertise.

In this paper we demonstrate the contributions of our pipeline, using a standalone optical sensor for both activity recognition and heart rate monitoring with significantly reduced sampling frequencies. This novel approach yields not only improved power efficiency but does so without significantly sacrificing accuracy thus advancing the development of simpler, more cost-effective wearables. 

Although globally people are using hospitals more efficiently, public healthcare expenditure is rising. For example, in Ireland expenditure
has risen from €14.9 billion in 2009 to an estimated €16.8 billion in 2018 with the increasing prevalence of chronic illness requiring long-term patient-provider engagement and management, accounting for roughly 80\% of spending \cite{Dept_Health2018, GPBullhound2015, Deloitte2019}. Frost \& Sullivan in 2010 predicted, based on the then-current trends, that healthcare spending in Western economies would almost double (as a proportion of GDP) by 2050, reaching 20\%-30\% of GDP in some cases. The report also stated that per capita, healthcare spending is rising faster than per capita income in most countries \cite{Frost2010}.

As a response, globally there is a change in how healthcare is managed. For example, in Ireland, the Department of Health has signalled a major shift in the paradigm of treating people with illnesses. This signals a change in health policy from a reactive to proactive treatment-based models where the focus is increasingly on keeping people healthy \cite{Dept_Health2012}. Advancements in digital health technologies, including mHealth and MedTech, have the potential to contribute significantly to a transformation in healthcare delivery, e.g. enabling proactive care through the use of continuous monitoring devices and application of advanced data analytics that enable greater personalisation of treatments \cite{Deloitte2019}. Thus the role of data gathered from wearables is important as part of such a shift in healthcare provision policies.

\section{Related Work}
Convolutional Neural Networks (CNN) have contributed tremendously to the success of machine learning since their introduction in the 1990s. They are an example of neuroscientific principles influencing deep learning \cite{Goodfellow2016}, in that they are designed to mimic the processing of images in the visual cortex of the human brain \cite{Raschka2017}. Fully automatic learning of a CNN allows the neural network to extract features that are salient in the input data across different layers. Given the correct training, a CNN allows for the implementation of high accuracy classifiers without the need for signal processing or feature extraction knowledge. This had contributed to their success in practical applications, particularly with image classification.

The current state of the art in HAR systems are camera-based which allow for direct capture of the data but consequently requires significant computer processing to determine distinct activities. HAR studies are frequently carried out using data from inertial measurement units (IMU) which measure proper acceleration of a body or limb.  Signal processing and feature extraction for these HAR studies are not trivial, including (but not limited to); singular value decomposition (SVD), support vector machine (SVM) and Random Forest (RF). High accuracy ranging from 80\% to 99\% can be achieved with such signal processing techniques but they often require a combination of sensor modalities \cite{Mannini2010} and using multiple IMUs located on various parts of the body which in turn gives rise to scalability and functionality issues in these studies.

Few studies have employed the use of a PPG sensor only for HAR as they are more commonly used with heart rate estimation \cite{Brophy2018, Boukhechba2019}. Biagetti {\em et al.} conducted a study on the same dataset used in this paper for activity recognition \cite{Biagetti2018}. Using the PPG data only for HAR they achieved 44.7\% classification accuracy using their feature extraction algorithm.  Later the authors combined the PPG data with accelerometer data and achieved 78.0\% accuracy using their feature extraction technique. Mehrang {\em et al.} used a combination of PPG and accelerometer with feature extraction and classification techniques such as RF and SVM, achieving accuracy of \SI{89.2 \pm 4.2}\% and \SI{85. \pm 6.8}\% respectively\cite{Mehrang2018}.

It should be noted that leading, modern feature extraction and classification techniques using multiple IMUs can achieve 80\% to 99\% classification accuracy, which may require several sensors located throughout the body \cite{Mannini2010}.

We have found few works that leverage CNN for heart rate estimation. Qui et. al computed heart rate estimation from facial videos using a CNN\cite{Qui2018}. In \cite{Spetlik2018}, the authors proposed a method to estimate heart rate using a CNN trained on a sequence of facial images. Reiss et. al sought to solve a regression problem by estimating heart rate from PPG and accelerometer data by computing the Fast Fourier Transform (FFT) and z-normalisation on the 4 input channels to a CNN \cite{Reiss2019}. Extending on this, using a standalone PPG we develop a CNN regression architecture for heart rate estimation on a single channel time series without any preprocessing.

Junker, Lukowicz and Tr\"{o}ster \cite{Junker2004} downsampled wearable accelerometers from 100 Hz in a wearable context recognition system. The authors found that they could achieve sufficient classification accuracy rates for sampling frequencies as low as 20 Hz. However, a significant drop in accuracy (below 60\%) is observed when the sampling rate is reduced to 10Hz.

Finally, in \cite{Krause2005} the authors use a developed wrist-worn wearable consisting of a two-axis accelerometer, microphone, light and temperature sensors for context-aware wearable computing. They found that a sampling frequency of 6 Hz yields comparative accuracy compared to much high sampling rates using available time domain features with machine learning. Following this Krause {\em et al.} demonstrated that this decrease in sampling frequency from 20Hz to 6Hz increases the battery life of their constructed wearable by 85\%.

\section{Methods}
\subsection{Computing Platform}
The experiments for this project were run on an Nvidia Titan Xp with Tensorflow. The code for these experiments are available online\footnote{GitHub Repository: https://github.com/Brophy-E/CNNs\_HAR\_and\_HR}.  

\subsection{Dataset}
A readily available wrist PPG exercise dataset collected by  Jarchi and Casson \cite{Jarchi2017} and publicly available on PhysioNet was used for the experiments in this paper \cite{PhysioNet}. Data was collected during exercise by 8 healthy patients (5 male, 3 female) with a sampling frequency of 256 Hz. Data was gathered using a wrist-worn PPG sensor on board the Shimmer 3 GSR+ unit for an average recording time of 4 to 6 minutes with a maximum time of 10 minutes. Four exercises were performed; two on a stationary exercise bike and two on a treadmill. The exercises are broken down as follows; walk on a treadmill, run on a treadmill, high resistance exercise bike and low resistance exercise bike. No further filtering is applied to the PPG data for the treadmill exercises other than what the Shimmer unit provides on board. For the exercise bike recordings there was high frequency noise present which was filtered in MATLAB using a second order IIR Butterworth filter with a 15Hz cutoff frequency \cite{Jarchi2017}. 

To accurately evaluate the unfiltered PPG heart rate performance, we compare it with a concurrent ECG that was collected by the authors of the data gathering experiment described above. This will provide a ground truth against which to assess our heart rate estimation.

\subsection{Downsampling and Segmentation}
Prior to segmenting and plotting the PPG signal it was downsampled to a number of different sampling frequencies. The classifier was trained in Python using the full 256Hz sampling frequency, then retrained on the downsampled frequencies of 30Hz, 15Hz, 10Hz, 5Hz and 1Hz respectively.

Once the signal had been downsampled it was then segmented into smaller chunks. A simple rectangular windowing function was used to capture 8 seconds worth of data and step through the data in increments of 1 second.  

\subsection{Human Activity Recognition}
\begin{figure}[ht]
\centering
\includegraphics[width=0.47\textwidth]{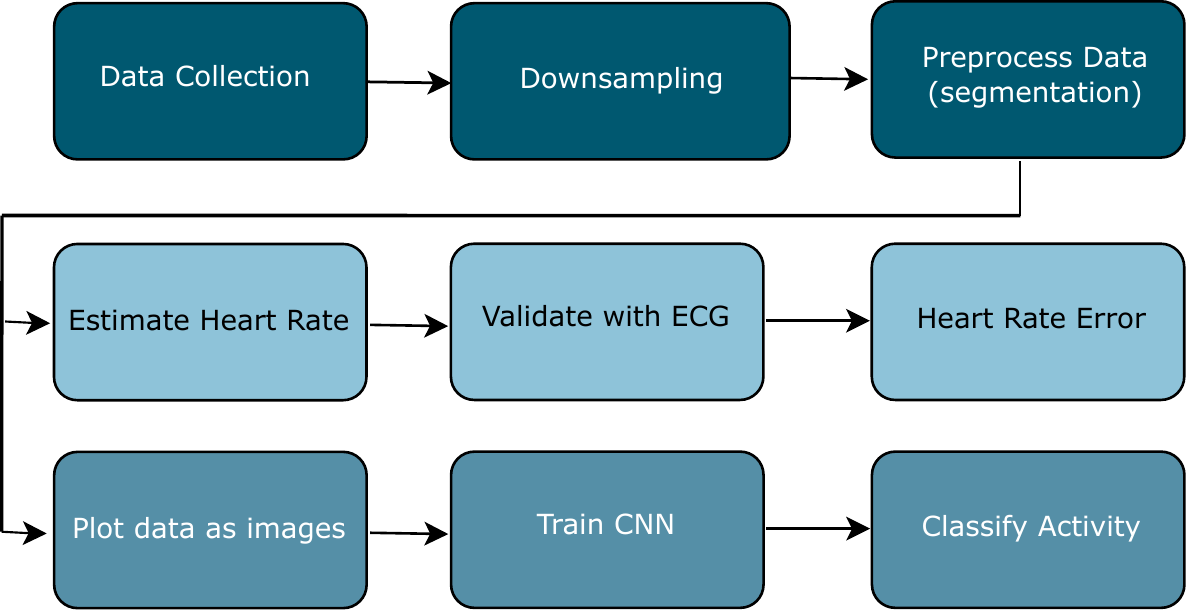}
\caption{Block diagram of our processing approach}
\label{fig:Block_Diagram}
\end{figure}

A CNN based on the Inception-V3 architecture and pre-trained on ImageNet was used as the classifier for the HAR experiments. The deep model was retrained leveraging the technique of transfer learning \cite{Pan2010}, the penultimate layer had its weights updated while all other layers remained the same. This allowed the use of smaller amounts of data to train a model with a large learning capacity that would normally require a lot of data and time to train from scratch. The retraining process can be fine-tuned through the optimisation of hyperparameters. The parameters were set as their defaults in this experiment except for the number of training steps which were changed from 10,000 to 4,000. This helped minimise overfitting through sufficient convergence of the loss function (cross-entropy). See Figure~\ref{fig:Block_Diagram} below for a block diagram of the processing pipeline associated with our methodology.

As a machine vision approach is applied using this classifier, the temporal PPG signals are saved as images rather than time series vectors to be used as input. Matplotlib, a Python plotting library was used to plot the PPG signal as images, which were saved as 299x299 JPEGs. All axis labels, legends, titles and grid ticks were removed. Python’s wfdb library was used to pull and load the data from PhysioNet.

\begin{figure}[ht]
\centering
\includegraphics[width=0.45\textwidth]{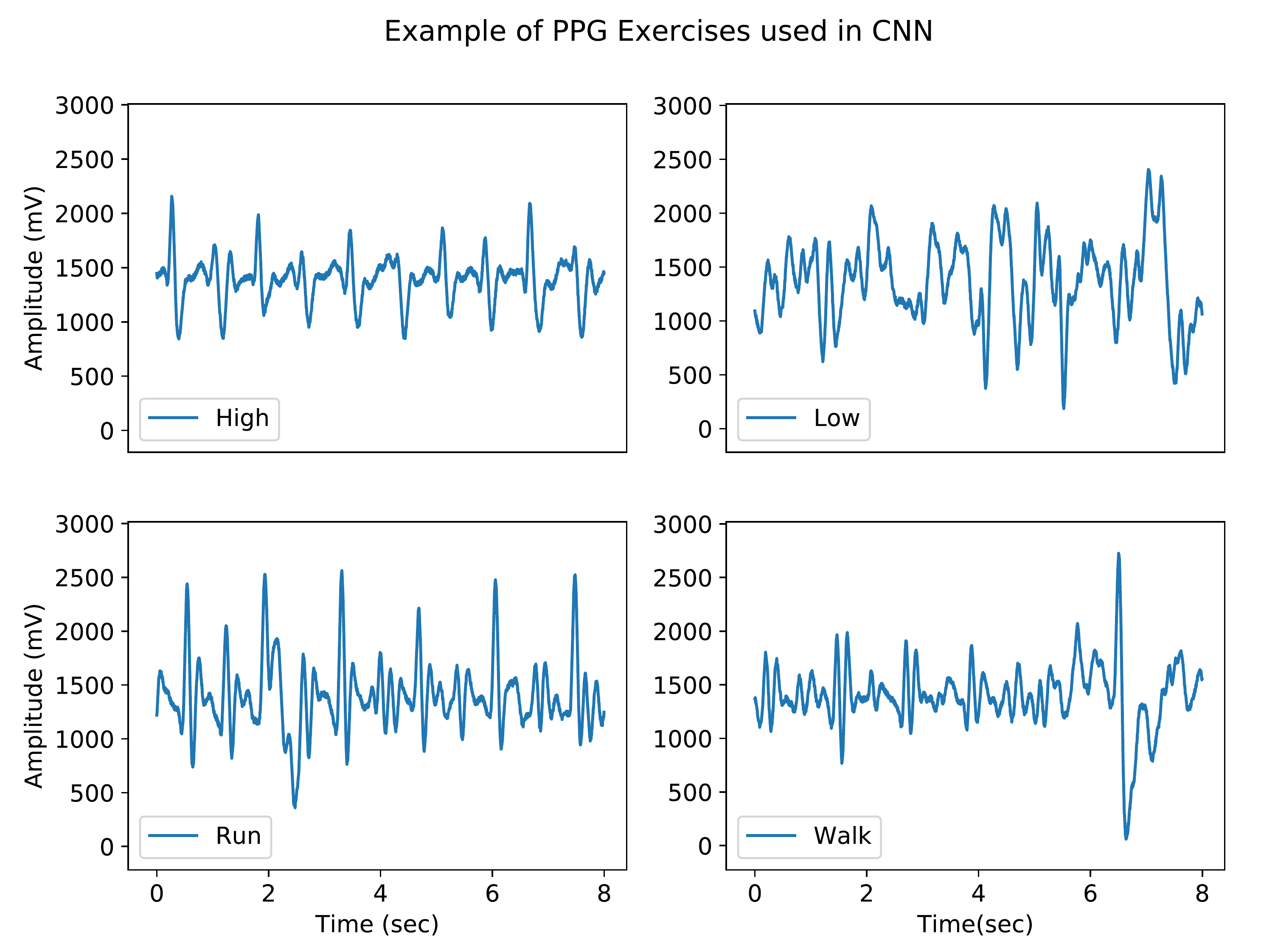}
\caption{Example of PPG from each exercise used in CNN training}
\label{fig:PPG_Exercises}
\end{figure}

To train the HAR classifier, a total of 6,653 images were stored in four sub-directories of the possible predicted classes (High, Low, Run and Walk). A train/test split validation approach was taken in this experiment. 80\% of the data was used for training, 10\% for validation and 10\% for testing. See Figure~\ref{fig:PPG_Exercises} for examples of PPG data used during training of the classifier.

\subsection{Estimation of Heart Rate}
We designed a CNN with the output layer replaced by a regression layer. We refer to this model as CNNR (Convolutional Neural Network with Regression). It is a four-layer 1-D convolutional network with batch normalization and ReLU (Rectified Linear Units) followed by a fully connected and regression layer respectively. The model architecture can be seen in Figure \ref{fig:CNNR_Arch} below. This model is used to estimate heart rate from the noisy PPG data. We used a train-test split of 90/10 for the CNNR.

\begin{figure}[ht]
\centering
\includegraphics[width=0.47\textwidth]{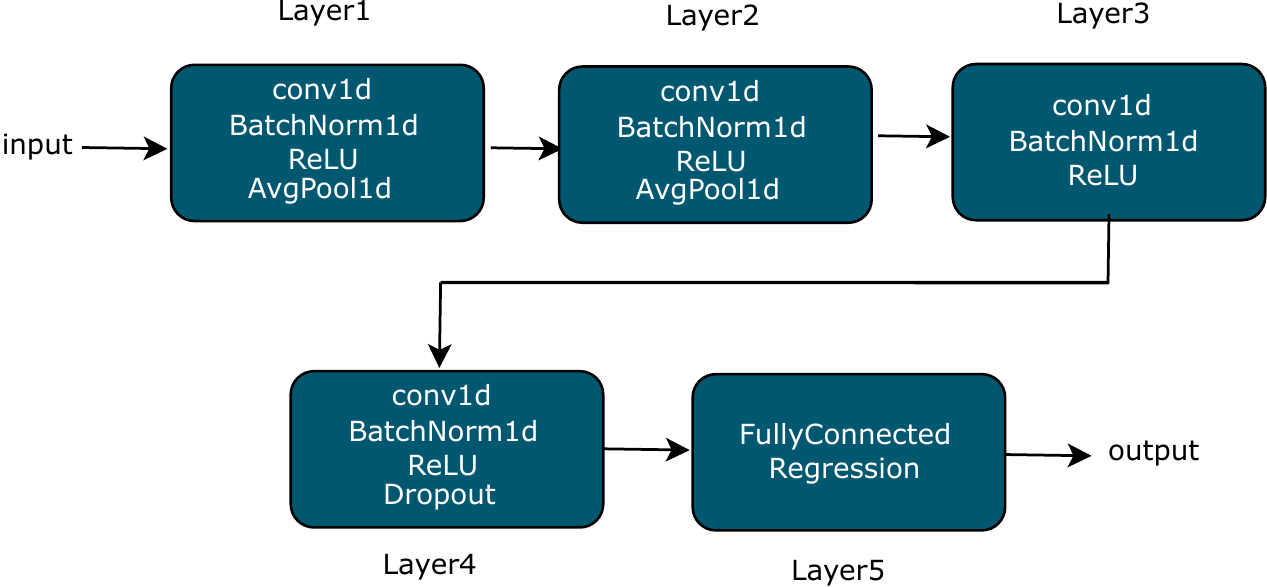}
\caption{Architecture of CNNR}
\label{fig:CNNR_Arch}
\end{figure}

HeartPy, an open-source toolkit for estimating heart rate from the PPG data, was used in our work as a baseline reference to compare the performance of our CNNR approach \cite{VanGent2018}. The HeartPy toolkit is designed to handle clean and noisy PPG data collected from either PPG or camera sensors. In the case of both our CNNR and HeartPy work, the PPG data used was the noisy, raw time-series signal. The estimated heart rate value for a segment of the signal was then compared to its concurrent ECG time series. The QRS peaks from the ECG were annotated as part of the data collection experiment. An estimated heart rate obtained using the CNNR and PPG toolkit on noisy data was then compared to the ECG heart rate which acted as the ground truth.

\section{Results}

\subsection{Human Activity Recognition}
The results for the HAR experiment are shown in Table~\ref{table:Sample_Freq_vs_Accuracy} below. As expected, the highest classification accuracy of 90.8\% is achieved when the original sampling frequency of 256Hz is used. However, we can still achieve a very competitive estimation performance even after downsampling the original sampling frequency to 5 Hz. Perhaps what is most surprising is the superior performance of our classifier when 10 Hz is chosen as the sampling frequency compared to the higher frequencies (15 Hz and 30 Hz) tested as part of this project. Due to the higher accuracy of 10Hz we also tested 12Hz, 11Hz, 9Hz and 8Hz as the chosen sampling frequency but found no anomaly as the surrounding frequencies yield similar accuracies. To further investigate the 10Hz performance, we low-pass filtered the PPG with a 4.5Hz cut-off frequency to remove possible aliasing but this did not impact the classification accuracy.

\begin{table}[ht]
\centering
\caption{Sampling Frequency vs. Accuracy}
\begin{tabular}{ |c|c|c| } 
\hline
Sample Frequency & Accuracy \\[0.5ex]
\hline \hline
256Hz		& 90.8\% \\
30Hz		& 82.3\% \\
15Hz		& 81.6\% \\
12Hz		& 82.1\% \\
11Hz		& 81.6\% \\
10Hz		& 83.0\% \\
9Hz		& 81.2\% \\
8Hz		& 80.5\% \\
5Hz		& 80.2\% \\
1Hz		& 68.5\% \\
\hline
\end{tabular}
\label{table:Sample_Freq_vs_Accuracy}
\end{table}

\begin{figure}[ht]
    \centering
    \includegraphics[width=0.45\textwidth]{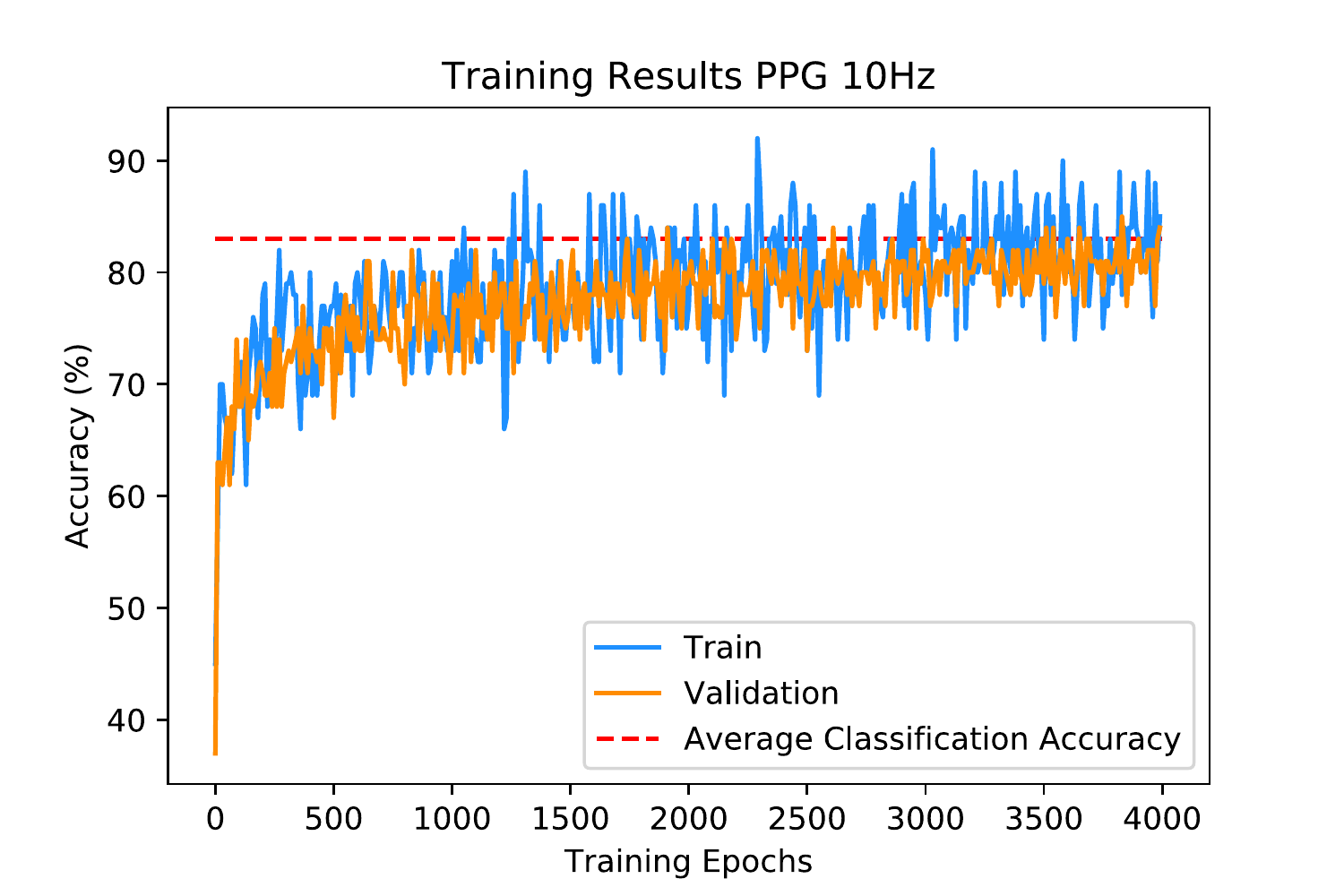}
    \caption{HAR Training Results for 10Hz Sampling Frequency}
    \label{fig:HAR_training_results_10Hz}
\end{figure}

\begin{figure}[hb]
    \centering
    \includegraphics[width=0.45\textwidth]{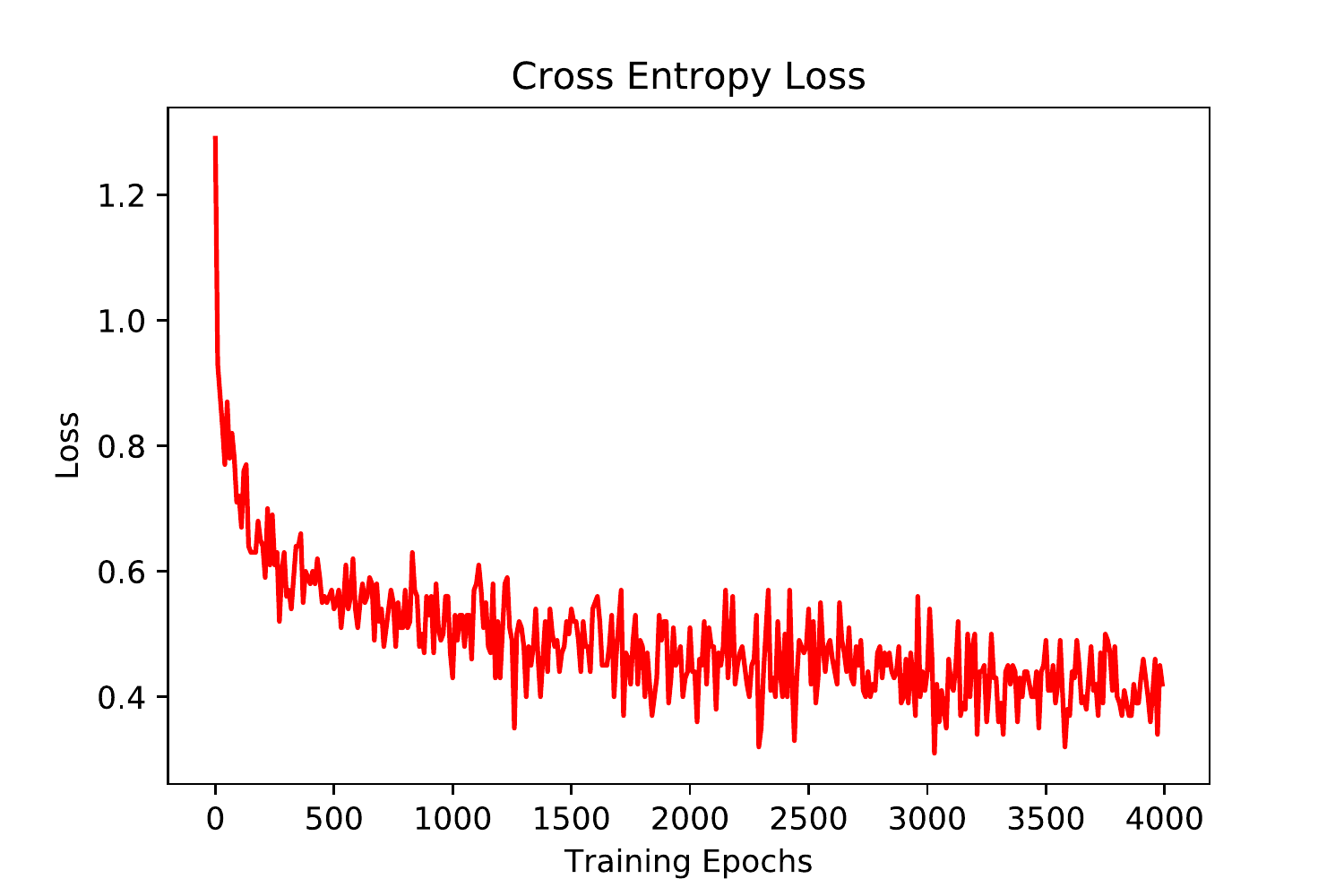}
    \caption{HAR Cross Entropy for 10Hz Sampling Frequency}
    \label{fig:HAR_cross_entropy_10Hz}
\end{figure}

\begin{figure}[ht]
    \centering
    \includegraphics[width=0.4\textwidth]{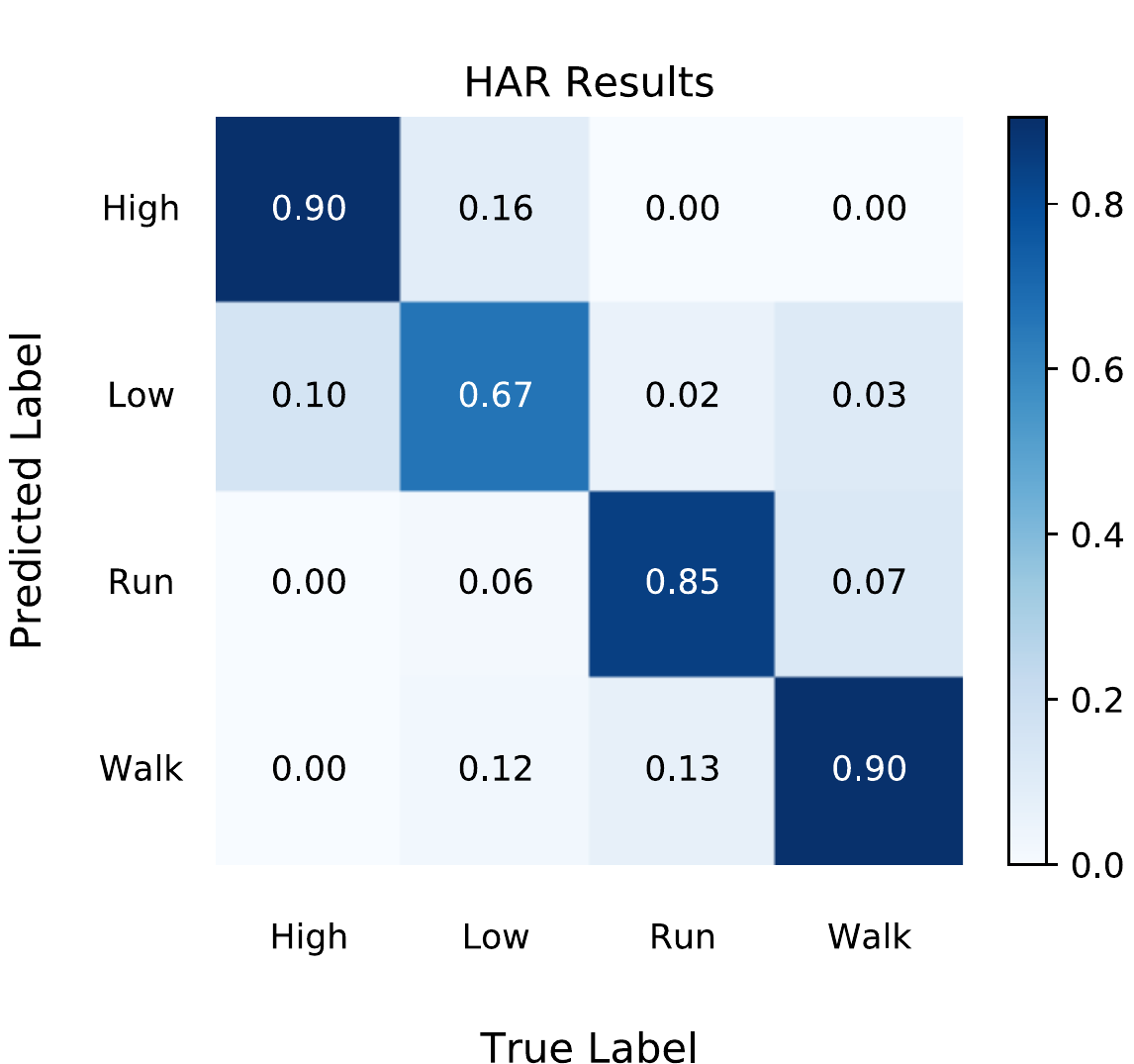}
    \caption{Confusion Matrix of HAR classifier}
    \label{fig:Confusion_Matrix}
\end{figure}

As a sampling frequency of 10 Hz performed the best out of the sampling frequencies tested, we show the training results for this sampling frequency over the 4,000 epochs along with the cross-entropy loss function and confusion matrix for exercise classification in Figures~\ref{fig:HAR_training_results_10Hz}, \ref{fig:HAR_cross_entropy_10Hz} and \ref{fig:Confusion_Matrix} respectively. We also show the relevant precision, recall and F1-scores in Table \ref{table:Precision_Recall}.

\begin{table}[ht]
\centering
\caption{Precision, Recall and F1-score}
\begin{tabular}{ |c|c|c|c| } 
\hline
Exercise & Precision & Recall & F1-Score\\[0.5ex]
\hline \hline
High	& 0.803	& 0.904 & 0.851 \\
Low		& 0.846 & 0.666 & 0.745 \\
Run		& 0.826 & 0.849 & 0.837 \\
Walk    & 0.834 & 0.899 & 0.865 \\
\hline
\end{tabular}
\label{table:Precision_Recall}
\end{table}

\subsection{Estimation of Heart Rate}
\begin{figure}[ht]
\centering
\includegraphics[width=0.47\textwidth]{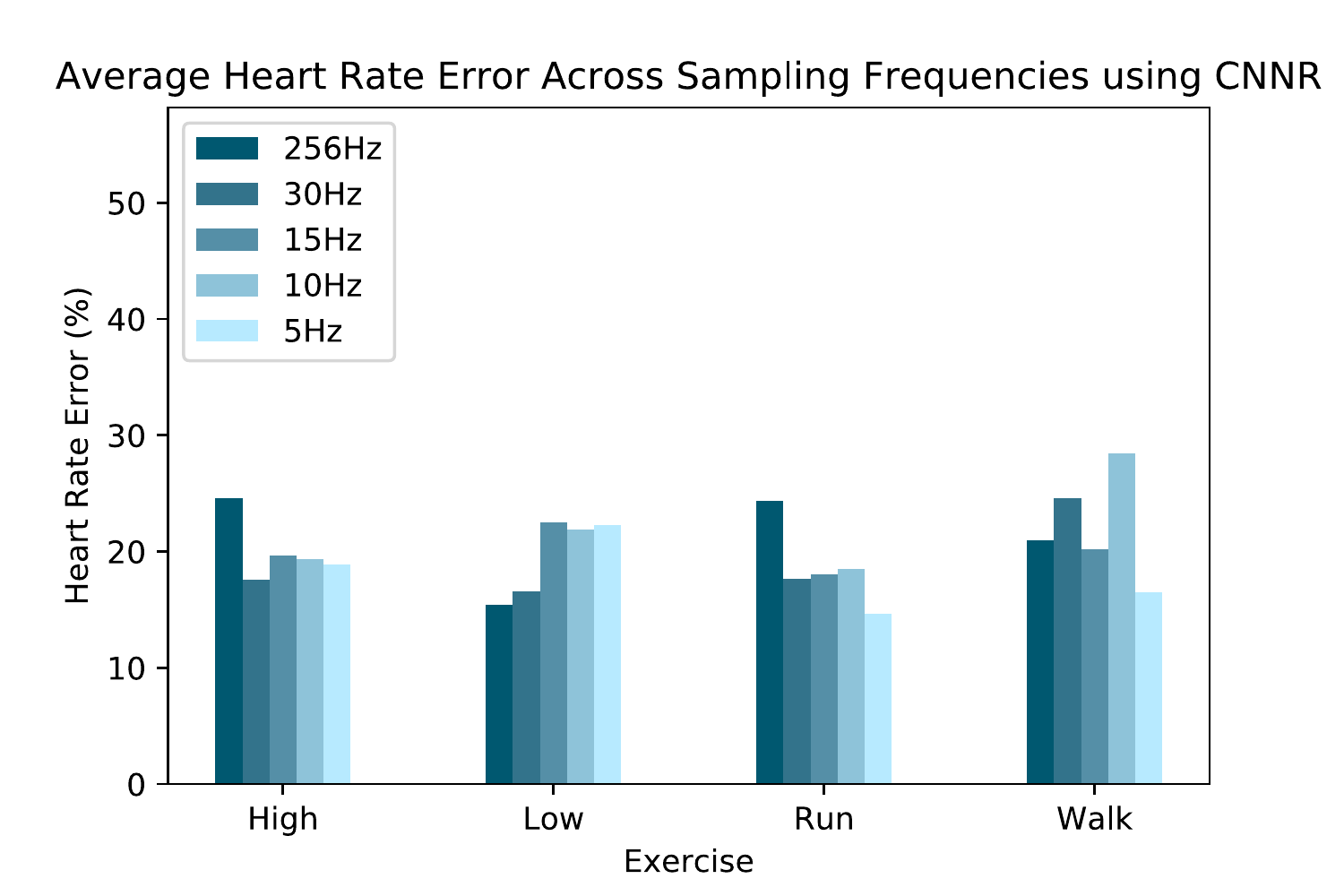}
\caption{Average Heart Rate Error Across all Exercises and Sampling Frequencies using our CNNR Method}
\label{fig:Heart_Rate_Error_CNNR}
\end{figure}

Results for estimating heart rate from the motion artefact (MA) corrupted PPG signal using HeartPy and our CNNR method are displayed below. Figure \ref{fig:Heart_Rate_Error} and Figure \ref{fig:Heart_Rate_Error_CNNR} presents the average heart rate error across the various sampling frequencies for each exercise for the two methods. The Heart Rate Error (HRE) is defined here as the absolute difference between the estimated heart rate for a given PPG sample and the heart rate ground truth calculated from the concurrent ECG sample.

\begin{figure}[ht]
\centering
\includegraphics[width=0.47\textwidth]{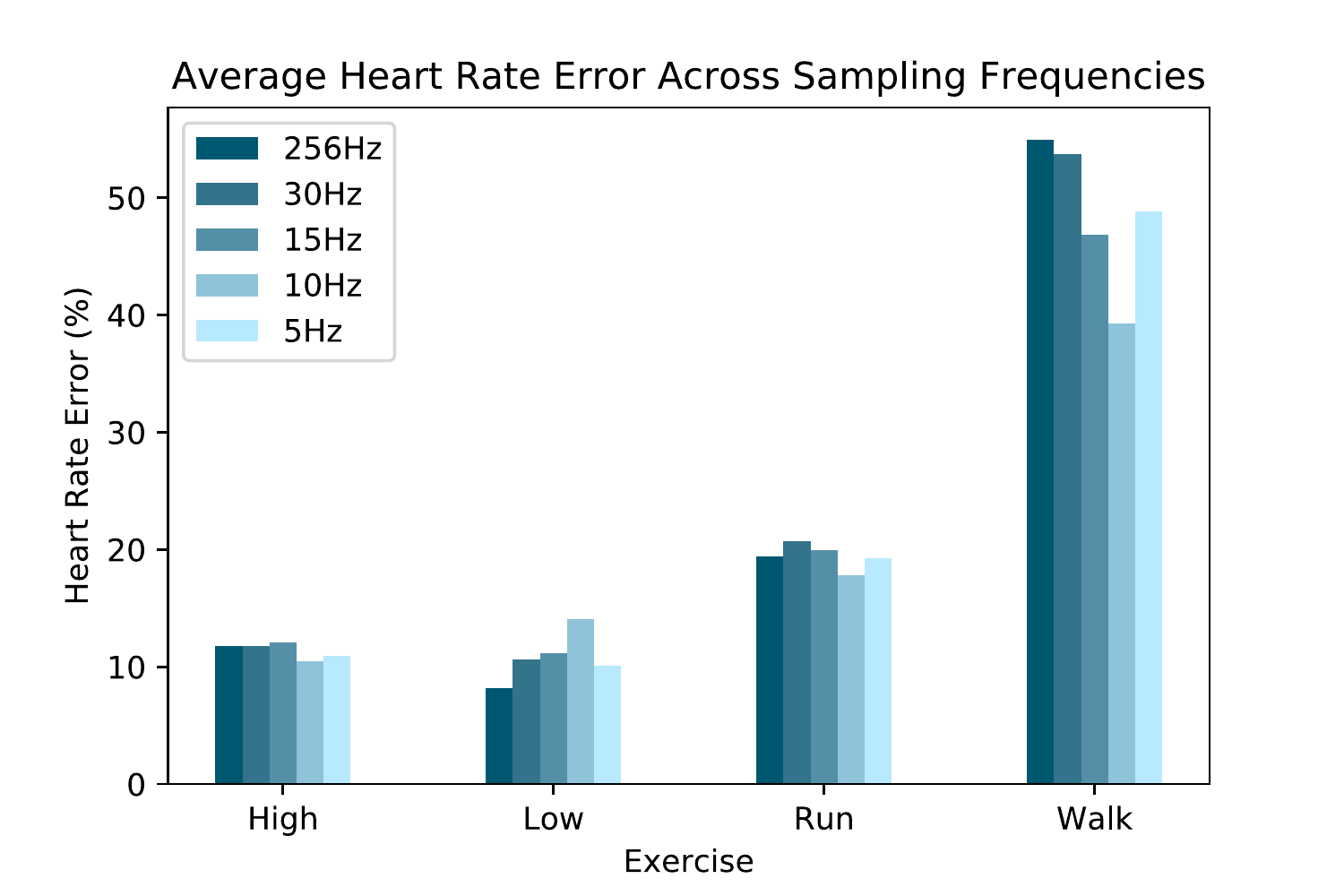}
\caption{Average Heart Rate Error Across all Exercises and Sampling Frequencies using HeartPy}
\label{fig:Heart_Rate_Error}
\end{figure}

For the HeartPy method, exercise specific HRE is similar across all sampling rates except from the 10 Hz sampling frequency on the walk exercise. Other sampling frequencies return an error of between 46\% and 55\% whereas the 10 Hz sampling frequency reduces the error to 39\%. The numerical results for the heart rate experiments is displayed in Table~\ref{table:HRE_HeartPy} where it can be clearly seen that 10Hz sampling frequency performs best for estimating heart rate from the MA corrupted signal.

\begin{table}[ht]
\caption{Heart Rate Error using HeartPy}
\begin{center}
\begin{tabular}{|c||c|c|c|c|}
    \hline
    \multirow{2}{*}{Sampling Frequency} & \multicolumn{4}{|c|}{Exercise}\\\cline{2-5}
    &High & Low & Run & Walk\\
    \hline \hline
    256Hz   & 11.78 & \textbf{8.14} & 19.44 & 54.94\\
    30Hz & 11.80 & 10.61 & 20.71 & 53.69\\
    15Hz  & 12.10 & 11.15 & 19.94 & 46.83\\
    10Hz & \textbf{10.46} & 14.05 & \textbf{17.82} & \textbf{39.28}\\
    5Hz & 10.94 & 10.05 & 19.27 & 48.85\\
    \hline
\end{tabular}
\end{center}
\label{table:HRE_HeartPy}
\end{table}

Our CNNR results can be found in Table \ref{table:HRE_CNNR} below. It can be seen that the HRE is similar across all exercises and there is not a distinguishable loss in accuracy for any of the sampling frequencies. For the walk exercise there is a great improvement in accurately estimating heart rate compared to the HeartPy method. It should be noted that average HRE across all exercises and sampling frequencies has decreased using the CNNR method from 22.59\% to 20.15\%, an increase in over 2 percentage points.

\begin{table}[!ht]
\caption{Heart Rate Error using our CNNR method}
\begin{center}
\begin{tabular}{|c||c|c|c|c|}
    \hline
    \multirow{2}{*}{Sampling Frequency} & \multicolumn{4}{|c|}{Exercise}\\\cline{2-5}
    &High & Low & Run & Walk\\
    \hline \hline
    256Hz   & 24.55 &  \textbf{15.41} & 24.38 & 20.94\\
    30Hz & \textbf{17.59} & 16.58  & 17.68 & 24.56\\
    15Hz  & 19.62  & 22.47  & 18.00 & 20.21\\
    10Hz & 19.30 & 21.87 & 18.49 & 28.41\\
    5Hz & 18.90 & 22.26 & \textbf{14.61} & \textbf{16.47}\\
    \hline
\end{tabular}
\end{center}
\label{table:HRE_CNNR}
\end{table}
 
\subsection{Optimisation of CNNR}
Following on from our results produced in \cite{Brophy2020}, we wanted to further decrease the heart rate error. Computing a non-exhaustive grid search over some of the hyperparameters for the CNNR returned an average HRE of 13.62\%, a decrease of nearly 7 percentage points from that of the CNNR without optimisation. We chose, number of epochs, learning rate and the train-test split as some parameters to optimise. The results for the optimisation process have been graphically presented in Figure \ref{fig:HRE_Optimised} below. To the authors knowledge, this is the best result using CNNs adapted for regression to estimate heart rate data from raw, noisy PPG sensor data. 
 
\begin{figure}[ht]
\centering
\includegraphics[width=0.47\textwidth]{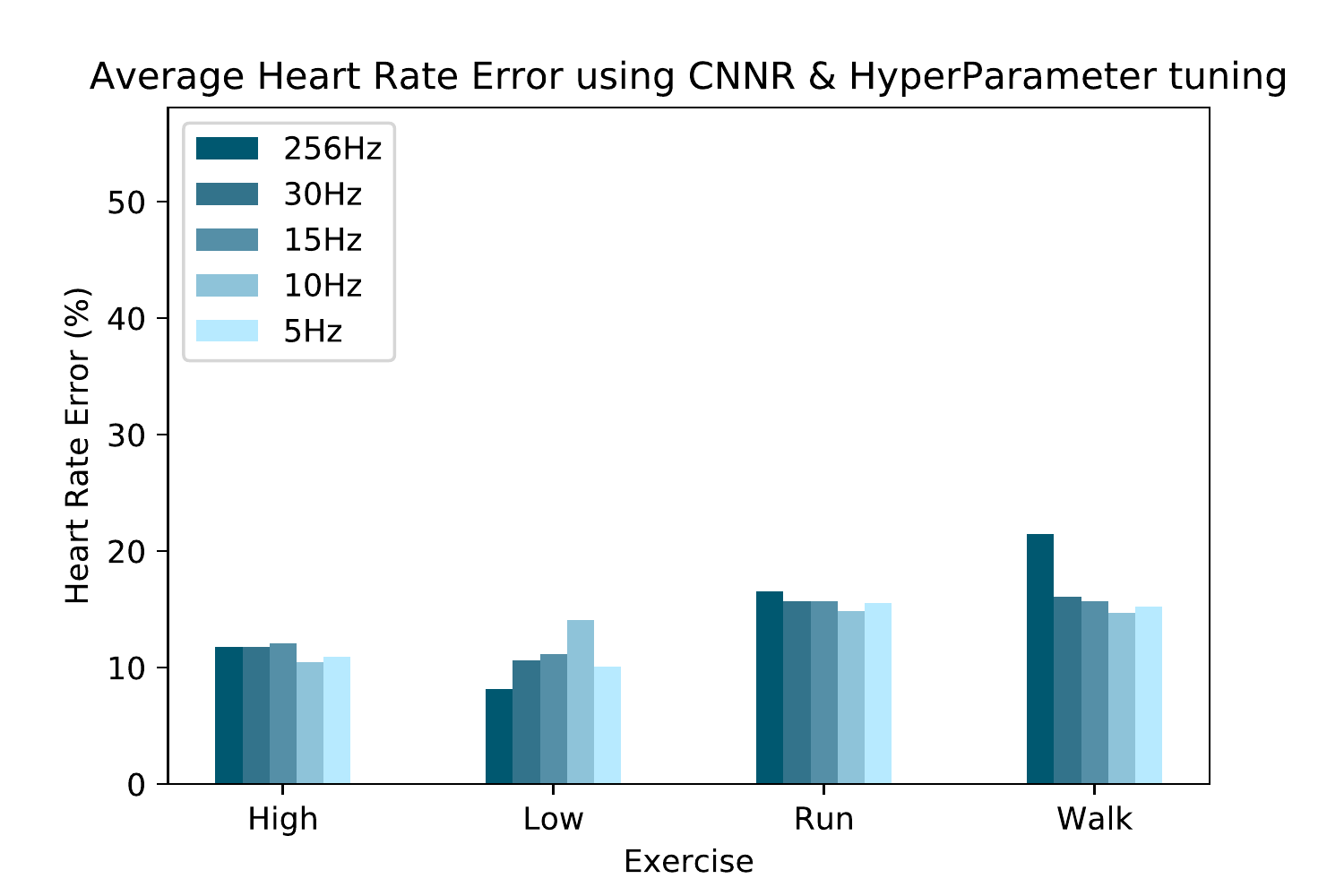}
\caption{Heart Rate Error using our Optimised CNNR}
\label{fig:HRE_Optimised}
\end{figure}

\section{Conclusion}
The approaches used in this paper yield highly competitive results for HAR even though only the optical signal is used.  This demonstrates that more cost and power-efficient wearables are possible through the exploitation of secondary information available from a simple optical sensor. This suggests single-sensor based wearables can achieve much of the functionality and capabilities of more complex multi-modal wearables.

The sampling rate did not have too much of an adverse effect on the performance of the algorithms. Interestingly, the CNN performed better at a 10 Hz sampling frequency compared to 15 Hz and 30 Hz. The reasons behind this have not been fully investigated.

Perhaps what was the most surprising from the results presented in this paper was the heart rate estimation results. We demonstrate how a CNN regression approach is capable of robust heart rate estimation even during periods of high artefact. The performance, particularly during these high artefact scenarios, was superior to conventional signal processing approaches for such estimation as demonstrated by the relative performance of the open-source tool kit HeartPy which served as a baseline here.  Furthermore, this heart rate estimation performance was sustained even at reduced sampling frequencies.  Notably sampling the sensor at 5 samples per second just as well as all other sampling frequencies, including the original 256 Hz.

A pervasive computing approach to wearables is taken here. Using a low power wearable with a single optical sensor and a sampling frequency of 10 Hz we can demonstrate compelling performance both in heart rate estimation and human activity recognition. This has the potential to reduce costs, improve battery performance and encourage wider adoption of digital technologies to a larger population and allow the transition to personalised, patient-centred preventative models of healthcare. Increasing access and affordability to these technologies will in turn lower costs and the strain on public healthcare expenditure, as well as helping to improve overall wellness.

\section*{Acknowledgment}

We gratefully acknowledge the support of NVIDIA Corporation with the donation of the Titan Xp used for this research.

\bibliographystyle{IEEEtran}
\bibliography{references}

\vspace{12pt}

\end{document}